\documentclass[a4paper,11pt]{article}
\usepackage{style/jheppub}
\usepackage[T1]{fontenc}
\usepackage{lmodern}
\usepackage{amsthm}
\usepackage{comment}
\usepackage{bm}
\usepackage{booktabs}
\usepackage{needspace}
\usepackage{placeins}
\allowdisplaybreaks[2]

\makeatletter
\newenvironment{sheetasymptotics}{%
    \par\begingroup\fontsize{8}{10}\selectfont
    \@fleqntrue\@mathmargin=0pt\relax
    \setlength{\jot}{3pt}%
}{\par\endgroup}
\makeatother
\newcommand{\sheetlabel}[1]{\makebox[4em][l]{$#1$}}

\title{Loop Equations for Multi-Matrix Models}

\author[a]{Aravinth Kulanthaivelu}

\affiliation[a]{Independent Researcher}

\emailAdd{aravinth.kulanthaivelu@outlook.com}

\abstract{%
    We revisit the combinatorial loop-equation method for computing planar disc amplitudes in
multi-matrix models with polynomial potentials. We develop a procedure which organises the Schwinger-Dyson
constraints by the level of their boundary insertions and, when the resulting system closes,
produces an algebraic equation for the planar disc function. Using this method we recover the disc
partition functions associated with fixed, mixed, and free boundary conditions in the three-state
Potts model, previously obtained by saddle-point methods. As a novelty we derive an algebraic equation for the fixed-boundary resolvent of the three-state Potts model with unequal quadratic and cubic couplings. We further show one can compute the disc
partition function for several unsolved two- and three-matrix models. These examples demonstrate that
loop equations provide a practical route to identifying and solving sectors of
multi-matrix models beyond the standard one- and two-matrix cases.
}

\keywords{matrix model, Potts model, boundary condition, triangulation}

\newcommand{\beq}{\begin{equation}}
\newcommand{\eeq}{\end{equation}}
\newcommand{\eref}[1]{(\ref{#1})}

\newcommand{\Tr}{\mathop{\mathrm{Tr}}\nolimits}

\newcommand{\rmd}{\mathrm{d}}

\theoremstyle{remark}

\theoremstyle{definition}

\begin{document}
\maketitle

\section{Overview}

Random matrix models are valuable tools for graph enumeration. Canonically, these formal matrix
integrals are identified with a sum over graphs. Each ribbon graph carries a factor $N^\chi$, where $\chi$ is the Euler characteristic and $N$ is the matrix size. This topological expansion is particularly relevant in the large $N$ limit
where planar graphs dominate. They have long been of interest in theoretical physics, where the
sum over graphs can be interpreted as a discretised path integral for two-dimensional Euclidean
quantum gravity. With multi-matrix models one is able to endow the graphs with statistical lattice
models, and hence study quantum gravity coupled to matter, or equivalently, bosonic string theory
in a non-critical target space \cite{DiFrancesco:1993cyw, ginsparg1993lectures, Ambjorn:1997di}.

In this article, we consider the class of matrix integrals of the form \beq Z = \int \rmd X_0 \rmd
    X_1 \ldots e^{-N \Tr V(X_0,X_1,\ldots)}
\eeq
where the $X_i$ are Hermitian $N\times N$ matrices with the flat Lebesgue measure on Hermitian matrices, invariant under unitary conjugation, and the
potential is a polynomial function. One is typically interested in evaluating the moments of the
model, which can be captured by a generating function known as the resolvent. We are particularly
interested in when the resolvent satisfies an algebraic equation in the planar limit, which we will
regard as ‘solvable’. The usual methods of solution can be broadly grouped into saddle-point
methods, orthogonal polynomials, and the method of loop equations. There is, however, no
straightforward mechanism for knowing if or when a multi-matrix model will be susceptible to such
techniques, and methods of solution are often very specific to the model under study.

Building on the work of \cite{Carroll:1995nj, kulanthaivelu2019freevariableloopequations} we revisit the elementary method of loop equations, also known as Schwinger-Dyson equations,
for multi-matrix models. We present a simple method for generating and solving the equations and
demonstrate its utility by applying the technique to several cases in the literature. We show that
several of the ‘unsolvable’ models studied using the bootstrap method
\cite{khalkhali2025bootstrappingcriticalbehaviormultimatrix} are indeed solvable. Moreover, we
demonstrate that the boundary conditions corresponding to sums of random matrices in the 3-state
Potts model, studied by Atkin, Niedner, and Wheater \cite{Atkin:2015ksy}, can also be derived via
this method, and we extend it to the 3-state Potts
model with unequal cubic potentials.

This article is organised as follows: in Section 2 we establish the formalism we use to manipulate
loop equations for general multi-matrix models. In Section 3 we describe an algorithmic approach
for computing and systematically solving loop equations. We immediately apply this in Section 4 for
the simple example of the two-matrix model with a general cubic potential that includes
cross-terms. Following this we turn to a set of two-matrix models studied in
\cite{khalkhali2025bootstrappingcriticalbehaviormultimatrix} by the bootstrap method
\cite{Lin_2020}. We then turn to the Potts model family and discuss the case of the fixed, mixed,
and free boundary conditions of the 3-state Potts model. Finally we provide a solution for the 3-state Potts model where the $S_3$
symmetry amongst spins is lifted. We provide code for the spectral curve calculations in \cite{Kulanthaivelu:LoopEquationsCode}.

\section{Preliminaries}
Consider a generic multi-matrix model in $n$ Hermitian matrix degrees of freedom
$\mathbf{X}=\{X_i\}_{i=0}^{n-1}$ of the form
\begin{equation}
    \label{loops1}
    Z = \int \prod_{i=0}^{n-1} \rmd X_i \, e^{-N\Tr V(\mathbf{X})},
\end{equation}
where $V(\mathbf{X})$ is a potential function which is a polynomial of degree $d+1$ in the $n$
matrices of $\mathbf{X}$, and depends on couplings $\{t_i\}_{i=2}^{\infty}$. We are interested in
computing the planar resolvent of a given matrix, say $X = X_0$, \beq\label{res} W(z) =
    \frac{1}{N}\left\langle \Tr \frac{1}{z-X}\right\rangle
\eeq
\noindent where here and in what follows $\langle\cdot\rangle$ denotes the average with respect to the measure \eref{loops1} and $z$ denotes the fugacity of a boundary link.

Define the following resolvent functions with boundary insertions
\beq\label{loopRes}
    W_{w(\mathbf{X})}(z) = \frac{1}{N} \left\langle \Tr  w(\mathbf{X}) \frac{1}{z-X} \right\rangle,
\eeq
where $w(\mathbf{X})$ is a word in the free algebra generated by $\mathbf{X}$, but with the
constraint that $X$ is not the first or last letter of the word. On the left hand side we will
proceed to label the resolvent functions by the indices labelling the matrices that comprise the
word $w(\mathbf{X})$ for the sake of brevity. Hence, one may write
\beq\label{resExample}
    W_{(1,0,2)}(z) = \frac{1}{N} \left\langle \Tr X_1 X_0 X_2\frac{1}{z-X} \right\rangle.
\eeq
Furthermore, we define the following trace correlation functions analogously,
\begin{equation}
    \label{trace}
    p_{w(\mathbf{X})}= \frac{1}{N} \left\langle \Tr w(\mathbf{X})\right\rangle.
\end{equation}
Define the level of a resolvent function of the form \eref{loopRes} as the length of the associated
word, $L(w(\mathbf{X}))$, such that, for example, \eref{resExample} with $L(X_1 X_0 X_2)=3$ denotes
a level-3 resolvent function. Similarly, we index trace correlation functions of a given level by
the set of words of appropriate length.

The number of inequivalent trace correlation functions of a given level is generally much smaller
than the size of the corresponding set of words. Cyclicity of the trace gives a $C_k$ symmetry. All
potentials considered here are also invariant under simultaneous transposition $X_i\mapsto X_i^T$,
which identifies a word with its reversal in expectation. Together these give the dihedral symmetry
$D_k$, of order $2k$. There may also be a global symmetry, $G$, of the matrix model potential,
given by
\beq
    \label{symmetry}
    G = \{ g\in \mathrm{Aut}(\mathbf{X}) : V(g(\mathbf{X}))=V(\mathbf{X})\}.
\eeq
An example of this would be permutation symmetry amongst the matrix degrees of freedom. Therefore,
defining the set of irreducible trace correlation functions of level $k$ as $\mathcal{P}_k$, we may
write
\begin{equation}
    \label{irreducibleTrace}
    \mathcal{P}_k \cong \mathbf{X}^k / (D_k \times G).
\end{equation}
The number of inequivalent correlators is then given by the cardinality of this set.

Similarly we define the set of irreducible resolvent functions of level $k$ as $\mathcal{W}_k$. The
number of independent resolvent functions is also constrained, although not as severely as the
correlation functions are. Define the operator $\Delta$, acting on resolvent functions as follows:
\begin{eqnarray}
    \label{delta}
    \Delta^k W_{w(\mathbf{X})}(z) & : = & \frac{1}{N} \left\langle \Tr  X^k w(\mathbf{X}) \frac{1}{z-X}\right\rangle \\
                                  & =   & z^k W_{w(\mathbf{X})}(z) - \sum_{i=1}^k z^{k-i} p_{X^{i-1}w(\mathbf{X})}.
\end{eqnarray}
We may omit words for which any sub-word consisting only of the matrix $X$ adjoins the loop
operator $(z-X)^{-1}$, since these may be expressed in terms of the $\Delta$ operator to a power
given by the length of the adjoining sub-word in $X$, acting on a lower-level resolvent function.
Moreover, the presence of the loop operator means that the dihedral symmetry is removed, but
reversal still gives $W_w(z)=W_{\operatorname{rev}(w)}(z)$. Finally, having fixed $X$ as the
representative matrix of the loop operator, the defining word only respects the subgroup
$G'\subseteq G$, consisting of elements $g\in\mathrm{Aut}(\tilde{\mathbf{X}})$, where
$\tilde{\mathbf{X}}=\mathbf{X} \setminus X$. Therefore we may write
\begin{equation}
    \label{irreducibleResolvent}
    \mathcal{W}_k \cong (\tilde{\mathbf{X}} \times \mathbf{X}^{k-2} \times \tilde{\mathbf{X}})/(\mathbb{Z}_2^{\mathrm{rev}}\times G'),\qquad k\geq2.
\end{equation}

For example, consider a three-matrix model with matrix degrees of freedom $X_0,X_1,X_2$, with a
$\mathbb{Z}_2$ symmetry under exchange of $X_0$ and $X_1$. Then for the resolvent functions with
respect to $X_1$, one would find $ W_{(2,1,0,1,2)} \neq W_{(2,0,1,0,2)}$, and therefore the
representative words comprise two distinct elements of $\mathcal{W}_5$, whereas the same words are
identified in $\mathcal{P}_5$ as one has $p_{2,1,0,1,2} = p_{2,0,1,0,2}$.

Loop equations are Schwinger-Dyson equations of the matrix model, expressing the invariance of the
matrix integral under infinitesimal reparameterisations $X_i \rightarrow X_i + \varepsilon \delta
    X_i$. There is some freedom as to which variations one can consider. For example, in the approach
of \cite{Eynard:2002kg}, specific to the two-matrix model, one computes reparameterisations using
products of the resolvent operators for the different matrices. The approach we adopt here is to
consider variations of the form
\begin{equation}
    \label{reparam}
    \delta X_i = w_1(\mathbf{X}) \frac{1}{z-X} w_2(\mathbf{X}).
\end{equation}
One then evaluates the loop equation by equating the first order contribution from the Jacobian
with the contribution from the potential. For a variation of $X=X_0$, differentiating the resolvent
gives the Jacobian contribution prescribed by the `split rule' \cite{Eynard:2002kg}, \beq
    \label{Jsplit} J_{\mathrm{res}}(X)=\Tr \left( w_1(\mathbf{X})
    \frac{1}{z-X}\right)\Tr\left(\frac{1}{z-X} w_2(\mathbf{X}) \right).
\eeq
There are additional Jacobian terms from differentiating any occurrences of the varied matrix in
$w_1$ and $w_2$. For variations of $X_i$ with $i\neq0$, only these inserted-word terms contribute,
since the resolvent is independent of $X_i$. In the large $N$ limit we assume the factorisation property $\frac{1}{N^2}\langle \Tr A \Tr B \rangle = \frac{1}{N}\langle \Tr A \rangle \frac{1}{N} \langle \Tr B \rangle + \mathcal{O}(N^{-2})$. Henceforth, all resolvent functions and correlators denote their planar limits. The resulting constraints may then be expressed in terms of
the resolvent functions \eref{loopRes} and trace correlation functions \eref{trace}.

We can classify loop equations by the level of their highest-level resolvent functions. If
$V'(\mathbf{X})$ has total degree $d$ in $\tilde{\mathbf{X}}$ then level-$k$ loop
equations are given by \eref{reparam} with $L(w_2)=k-d-L(w_1)$. We can therefore define a
corresponding set of level $k$ variations,
\begin{equation}
    \label{var}
    \mathcal{V}_k = \left\{ (w_1,w_2)\in\mathbf{V}_m\times\overline{\mathbf{V}}_{k-d-m} : m=k-d,\cdots, \left\lceil\frac{k-d}{2}\right\rceil \right\},
\end{equation}
where $\mathbf{V}_m=\mathbf{X}^{m-1}\times \tilde{\mathbf{X}}$ and $\overline{\mathbf{V}}_m =
    \tilde{\mathbf{X}}\times\mathbf{X}^{m-1}$, and we set $\mathbf{V}_0 = \overline{\mathbf{V}}_0 = \{1\}$, the empty word. That is, the words $w_1$ and $w_2$ are constrained such
that the letters in contact with the loop operator are in $\tilde{\mathbf{X}}$. This is so that the
resulting loop equations are distinct from lower-level loop equations.

The key idea of the loop equation approach we adopt is to compute the Schwinger-Dyson constraints
for each matrix degree of freedom for a sufficiently large set of variations. We aim to compute
loop equations up to a level $n_{\mathrm{max}}$ with enough independent constraints to eliminate
the auxiliary resolvent functions and retain an algebraic equation for the planar (level-0)
resolvent. Comparing the number of equations with the number of auxiliary resolvent functions to be
eliminated provides a useful guide to the required level. Of course this alone does not establish
closure as dependencies among the equations and relations among the correlators can change the effective rank. Instead, closure must be checked by the
elimination itself. This may be regarded as a discrete version of loop equation solution to the
two-matrix model \cite{Eynard:2002kg}, since one may expand the product of resolvent operators as
an infinite sum of words in matrices of arbitrary length multiplying a single resolvent operator.
It is hoped that this refined set of equations contains more information than if one only considers
a small set of reparameterisations by products of loop operators, allowing us to compute the planar
resolvent.

\section{General solution strategy}

In practice, for complicated multi-matrix models, one needs to consider a large number of loop
equations - on the order of hundreds for many of the highly symmetric three-matrix models we will
consider. Each of these encodes a combinatorial relation relating a random graph with prescribed
boundary conditions to a different set of random graphs with various boundary conditions when a
marked edge has been removed. Many of these are not independent, and so we need a systematic
approach for eliminating the higher-order resolvent functions.

For all $v\in \mathcal{V}_n$ we obtain a corresponding loop equation, which we denote
\begin{equation}
    \label{loopEq}
    \sum_{w\in\mathcal{W}_n} M_{v,w}^{(n)}\left(\{t_i\},\{p_i\}\right) \, W_{w}(z) + \mathcal{R}_v \big[ \bigcup_{n'<n} \mathcal{W}_{n'} \big] = 0,
\end{equation}
where $\mathcal{R}_v$ denotes the remaining terms in the loop equation containing resolvent
functions with level $m < n$. The coefficient matrix $M^{(n)}_{v,w}$ will generically depend on
resolvent functions of level $m \leqslant \lfloor n/2 \rfloor$, which has been suppressed in
\eref{loopEq}. Starting with \eref{loopEq} for level $n=n_{\mathrm{max}}$, one may identify a
linearly independent set of constraints with respect to the level-$n$ resolvent functions by using
the Gaussian elimination algorithm on $(M^{(n)})^T$. Gaussian elimination reduces $(M^{(n)})^T$ to
a matrix, $T^{(n)}$, in row echelon form. The rows of $(M^{(n)})_{ji}$ which form a linearly
independent basis are then the rows for which $j\in\mathcal{I}_n$ where
\begin{equation}
    \label{linInd}
    \mathcal{I}_n = \bigcup_i \left\{ \min\left\{ j : T_{ij}^{(n)} \neq 0 \right\} \right\},
\end{equation}
where the union is over the nonzero rows of $T^{(n)}$. One may then solve the linearly independent
set of equations to express the level-$n$ resolvent functions in terms of lower order resolvent
functions. Using these expressions to remove the level-$n$ resolvent functions from the remaining
constraints, one obtains a set of constraints linear in resolvent functions of top-level
$n_{\mathrm{max}}-1$, which may include distinct loop equations from those computed using
variations $v\in\mathcal{V}_{n-1}$. Together with \eref{loopEq} for
$v\in\mathcal{V}_{n_{\mathrm{max}}-1}$ we obtain an augmented set of relations at level
$n_{\mathrm{max}}-1$.

We iterate downwards while the highest-level resolvent functions enter linearly. Let $L_{\max}$ be
the maximum of $L(w_1)+L(w_2)$ over the retained variations of $X$. The quadratic Jacobian terms
$W_uW_v$ satisfy $L(u)+L(v)\leq L_{\max}$, so at least one factor has level at most
\begin{equation}
    r=\left\lfloor\frac{L_{\max}}2\right\rfloor.
\end{equation}
The system is therefore jointly linear in all resolvents above level $r$, treating the lower-level
resolvents as coefficients. Subject to full column rank and nonvanishing pivots, this upper sector
can be eliminated linearly. For a uniform hierarchy with potential-derivative degree $d$ and
$n_{\max}=L_{\max}+d$, the bound is $r=\lfloor(n_{\max}-d)/2\rfloor$. At level $r$ and below, the
remaining equations may be nonlinear and can be solved using a variety of techniques.

\begin{table}[htbp]
    \centering
    \begin{tabular}{@{}lrrr@{}}
        \toprule
        Spectral curve                           & $n_{\mathrm{max}}$ & $N_{\mathrm{res}}$ & $N_{\mathrm{cor}}$ \\
        \midrule
        Generic cubic two-matrix                 & 4                  & 10                 & 6 \\
        Quartic two-matrix $(g,g,g)$, $(g,-g,g)$ & 6                  & 11                 & 1 \\
        Quartic two-matrix $(g,g,-g)$, $(-g,g,g)$ & 7                  & 27                 & 4 \\
        Cubic three-matrix                       & 5                  & 64                 & 7 \\
        3-state Potts, fixed                     & 5                  & 64                 & 7 \\
        3-state Potts, mixed                     & 6                  & 245                & 17 \\
        3-state Potts, free                      & 6                  & 245                & 36 \\
        General 3-state Potts, fixed             & 6                  & 633                & 84 \\
        \bottomrule
    \end{tabular}
    \caption{Levels and equation counts in the accompanying notebook derivations.
        Here $n_{\mathrm{max}}$ is the largest resolvent-word level in the generated
        system, and $N_{\mathrm{res}}$ and $N_{\mathrm{cor}}$ count the nonzero
        resolvent and correlator equations after symmetry identifications
        and removal of duplicates, before elimination.}
    \label{tab:curve-counts}
\end{table}
\noindent Table~\ref{tab:curve-counts} records the size of the loop systems used below. For cubic potentials
the maximal variation-word length is $n_{\mathrm{max}}-2$, and for quartic potentials it is
$n_{\mathrm{max}}-3$. 

In the following we write each spectral curve as $F(x,y)=0$, with $F_{(p)}(x,y)=0$ for the boundary condition
associated with a sum of $p$ matrices. We display the associated asymptotic expansions for each curve in appendix~\ref{app:asympt}. Coefficients containing undetermined correlation functions
are denoted by $c_{i,j}$, where $i$ and $j$ are the exponents of $x$ and $y$. These coefficients
depend on the couplings and the correlators and exist purely
to simplify the presentation. The accompanying notebooks \cite{Kulanthaivelu:LoopEquationsCode} give the complete coefficient expressions and the calculations leading to each curve. Correlators left undetermined by the reduction parameterise a family of spectral curves. Selecting a particular solution requires additional analytic conditions, such as an appropriate one-cut condition. We will not develop this further in this article.

\section{Generic Cubic Two-Matrix Model}
To illustrate our formalism and demonstrate the strength of the loop equation approach, let us
consider the simple example of a two-matrix model with a generic cubic potential:
\begin{align}
    \label{2MMGeneric}
    V(M_1,M_2) = & \frac{1}{2(\alpha \beta - \gamma^2)}(\alpha M_1^2 + \beta M_2^2 - 2 \gamma M_1 M_2) \\
                 & \qquad \quad \nonumber - \frac{1}{3} (a M_1^3 + 3b M_1^2 M_2 + 3d M_1 M_2^2 + e M_2^3).
\end{align}
This furnishes an example of a multi-matrix model which is not solvable through any known
saddle-point or orthogonal polynomial approach. In particular, this appears to fall outside the domain of applicability of the loop equation method of \cite{Eynard:2002kg}. However, as first demonstrated in
\cite{Carroll:1995nj}, one may compute the resolvent of this model using our more granular loop
equation strategy.

For \eqref{2MMGeneric} the following are a sufficient set of reparameterisations:
\begin{align}
    \label{2MMGenericReps}
    \delta M_{i} & = \frac{1}{z-M_1}, \\
    \delta M_{i} & = M_2 \frac{1}{z-M_1} + \text{h.c.}, \\
    \delta M_{i} & = M_2^2 \frac{1}{z-M_1}+\text{h.c.}, \\
    \delta M_{i} & = M_2 \frac{1}{z-M_1} M_2,
\end{align}
for $i\in\{1,2\}$. Calculating the corresponding loop equations yields the following relations,
where the action of the $\Delta$ operator is given by \eqref{delta} for $k=1$, i.e.
$\Delta^1=\Delta$, and we suppress the $z$-dependence of the resolvent functions\footnote{Note that
    $W_2$ is distinct from the resolvent $W_{(2)}$ \eqref{discfct}, which will denote the resolvent of the
    sum of two matrices.}:
\begin{align}
    W^2 =             & -d W_{(2,2)}+\frac{\gamma  W_{2}}{\gamma ^2-\alpha  \beta }-a\Delta^2 W -2b \Delta W_{2} +\frac{\alpha \Delta W}{\alpha  \beta -\gamma ^2}, \\
    0 =               & -e W _{(2,2)}+\frac{\beta  W_{2}}{\alpha  \beta -\gamma ^2} - b\Delta^2 W  - 2d\Delta W_{2}+\frac{\gamma \Delta W}{\gamma ^2-\alpha  \beta }, \\
    W  W_{2} =        & \frac{\gamma  W_{(2,2)}}{\gamma ^2-\alpha  \beta }-a\Delta^2 W_{2} -b \Delta W_{(2,2)} -b W_{(2,1,2)}-d W_{(2,2,2)}+\frac{\alpha  \Delta W_{2}}{\alpha  \beta -\gamma ^2}, \\
    W =               & \frac{\beta  W_{(2,2)}}{\alpha  \beta -\gamma ^2}-b \Delta^2 W_{2}-d\Delta W_{(2,2)}-d W_{(2,1,2)}-e W_{(2,2,2)}+\frac{\gamma \Delta W_{2}}{\gamma ^2-\alpha  \beta }, \\
    W  W_{(2,2)} =    & \frac{\gamma  W_{(2,2,2)}}{\gamma ^2-\alpha  \beta }-a\Delta^2 W_{(2,2)}-b\Delta W_{(2,2,2)}-b W_{(2,1,2,2)} \\
                      & \qquad \qquad -d W_{(2,2,2,2)}+\frac{\alpha  \Delta W_{(2,2)}}{\alpha  \beta -\gamma ^2}, \nonumber \\
    W  p_{2}+ W_{2} = & \frac{\beta  W_{(2,2,2)}}{\alpha  \beta -\gamma ^2}-b\Delta^2 W_{(2,2)}-d\Delta W_{(2,2,2)}-d W_{(2,1,2,2)} \\
                      & \qquad \qquad -e W_{(2,2,2,2)}+\frac{\gamma \Delta W_{(2,2)}}{\gamma ^2-\alpha  \beta }, \nonumber \\
    W_{2}^2 =         & \frac{\alpha  W_{(2,1,2)}}{\alpha  \beta -\gamma ^2}+\frac{\gamma  W_{(2,2,2)}}{\gamma ^2-\alpha  \beta }-a W_{(2,1,1,2)}-2 b W_{(2,1,2,2)}-d W _{(2,2,2,2)}, \\
    2 W_{2} =         & \frac{\gamma  W_{(2,1,2)}}{\gamma ^2-\alpha  \beta }+\frac{\beta  W_{(2,2,2)}}{\alpha  \beta -\gamma ^2}-b W_{(2,1,1,2)}-2 d W_{(2,1,2,2)}-e W_{(2,2,2,2)}.
\end{align}
These are eight independent equations in eight unknowns which may be eliminated to give
$F(z,W(z))=0$, with $(\deg_x F,\deg_y F)=(5,4)$ for generic couplings.

\section{Quartic and cubic multi-matrix models}
\label{sec:multimatrix}

In this section we apply our method to the family of quartic two-matrix models and the three-matrix model considered by \cite{khalkhali2025bootstrappingcriticalbehaviormultimatrix}. There the authors adopted the matrix bootstrap which combined
loop equations with positivity constraints to bound the planar correlators of these models. They used these to estimate 
the critical couplings and the string susceptibility exponents.

Here we revisit these models using the loop-equation procedure of Section 3. We show that a finite set of loop equations can 
be reduced to an algebraic equation for each planar resolvent, providing an analytic description. The resulting spectral curves 
are presented below, while Appendix C provides a complementary treatment of the quartic models using standard matrix-model methods 
to confirm and determine the planar moments and critical behaviour.

\subsection{Quartic two-matrix models}
Consider the ordered trace potential
\begin{equation}
    V(A,B)=\frac12(A^2+B^2)+\frac{\alpha}{4}(A^4+B^4)
    +\frac{\beta}{2}ABAB+\gamma A^2B^2.
\end{equation}
For each of the following choices of couplings, the resolvent of $A$ obeys
\begin{equation}
    F\!\left(z,W_A(z)\right)=0.
\end{equation}
The coefficients $c_{i,j}$ refer to the curve in the corresponding case.

\subsubsection{$(\alpha,\beta,\gamma)=(g,g,g)$ or $(g,-g,g)$}
Both choices give the same form of the spectral curve,
\begin{align}
    \label{curve:quartic_ggg}
    F(x,y)={} & -y^{4}+y^{3}\left(gx^{3} + x\right)+y^{2}\left(gx^{2} -1-4g p_{0,0}\right)-gxy-g.
\end{align}

where $p_{0,0}=\frac{1}{N}\langle\Tr A^2\rangle$. The curve has degrees $(\deg_xF,\deg_yF)=(3,4)$.

\subsubsection{$(\alpha,\beta,\gamma)=(g,g,-g)$}
\begin{align}
    \label{curve:quartic_ggm}
    F(x,y)={} & -gy^{6}{}+y^{5}\left(g^2x^{3} + 16 gx\right)\notag \\
              & {}+y^{4}\left(-15 g^2x^{4} + (g-99) gx^{2} - 4 (g-3)\right)\notag \\
              & {}+y^{3}\left(84 g^2x^{5} - 8 (g-35) gx^{3} + c_{1,3}x\right)\notag \\
              & {}+y^{2}\left(-208 g^2x^{6} + 4 (g-83) gx^{4} + c_{2,2}x^{2} + c_{0,2}\right)\notag \\
              & {}+y\left(192 g^2x^{7} + 96 g (g+1)x^{5} + c_{3,1}x^{3} + c_{1,1}x\right)\notag \\
              & {}-192 g^2x^{6} + c_{4,0}x^{4} + c_{2,0}x^{2} + c_{0,0}.
\end{align}

The curve has degrees $(\deg_xF,\deg_yF)=(7,6)$.

\subsubsection{$(\alpha,\beta,\gamma)=(-g,g,g)$}
\begin{align}
    \label{curve:quartic_mgg}
    F(x,y)={} & gy^{6}{}+y^{5}\left(g^2x^{3} - 16 gx\right)\notag \\
              & {}+y^{4}\left(-15 g^2x^{4} + g (g+99)x^{2} + 4 (g+3)\right)\notag \\
              & {}+y^{3}\left(84 g^2x^{5} - 8 g (g+35)x^{3} + c_{1,3}x\right)\notag \\
              & {}+y^{2}\left(-208 g^2x^{6} + 4 g (g+83)x^{4} + c_{2,2}x^{2} + c_{0,2}\right)\notag \\
              & {}+y\left(192 g^2x^{7} + 96 (g-1) gx^{5} + c_{3,1}x^{3} + c_{1,1}x\right)\notag \\
              & {}-192 g^2x^{6} + c_{4,0}x^{4} + c_{2,0}x^{2} + c_{0,0}.
\end{align}

The curve again has degrees $(\deg_xF,\deg_yF)=(7,6)$. Remarkably we observe the same curve as \eqref{curve:quartic_ggm} under the map $g\rightarrow -g$.

\subsection{Cubic three-matrix model}
For the potential
\begin{equation}
    V(A_0,A_1,A_2)=\frac12\sum_{i=0}^2A_i^2
    +\frac g3\sum_{i=0}^2A_i^3+g(A_0A_1A_2+A_0A_2A_1),
\end{equation}
the planar resolvent satisfies $(\deg_xF,\deg_yF)=(6,7)$ with
\begin{align}
    \label{curve:cubic_three}
    F(x,y)={} & g^2y^{7}{}+y^{6}\left(-10 g^3x^{2} - 22 g^2x - 4 g\right)\notag \\
              & {}+y^{5}\left(9 g^4x^{4} + 138 g^3x^{3} + 187 g^2x^{2} - 2 g \left(5 g^2-17\right)x + c_{0,5}\right)\notag \\
              & {}+y^{4}\left(-108 g^4x^{5} - 594 g^3x^{4} + 6 g^2 \left(3 g^2-68\right)x^{3} + \sum_{i=0}^{2}c_{i,4}x^i\right)\notag \\
              & {}+y^{3}\left(324 g^4x^{6} + 540 g^3x^{5} - 18 g^2 \left(9 g^2-10\right)x^{4} + \sum_{i=0}^{3}c_{i,3}x^i\right)\notag \\
              & {}+y^{2}\left(324 g^4x^{5} + \sum_{i=0}^{4}c_{i,2}x^i\right)\notag \\
              & {}+y\left(-324 g^4x^{4} + \sum_{i=0}^{3}c_{i,1}x^i\right)\notag \\
              & {}-324 g^4x^{3} + \sum_{i=0}^{2}c_{i,0}x^i.
\end{align}

\section{Boundary conditions of the Potts model}

The $q$-state Potts matrix model was first proposed in \cite{Kazakov1988} wherein the solution was discussed for
the case of $q = 1$ and $q = 0$. For $q = 2$ we recover the Ising model on random surfaces, a
two-matrix model, which has been solved \cite{Kazakov1986, Boulatov1987, Eynard:2002kg}. For the
particular case of triangulations, corresponding to cubic potentials, and integer $0\leqslant q \leqslant 4$, the analytic
properties of the single-matrix resolvent were first computed by Daul \cite{Daul:1994qy} using
saddle-point methods. This was extended by Zinn-Justin \cite{ZinnJustin:1999jg}, who first defined
the dilute $q$-state Potts matrix model and was able to derive an algebraic expression for the
resolvent function. Using the method of loop equations Bonnet and Eynard \cite{Eynard:1999gp} were
able to demonstrate that the resolvent satisfies an algebraic equation for rational
$\arccos\!\left((q - 2)/2\right) / \pi $. Atkin, Niedner, and Wheater \cite{Atkin:2015ksy}
demonstrated that the resolvents for sums of random matrices also satisfy algebraic equations for
integer $0< q < 4$ using saddle-point methods.

It is common to regard resolvents as describing boundary conditions on random graphs. In the
diagrammatic expansion of the matrix integral each monomial of order $k+2>2$ generates a
$(k+2)$-sided polygon. To leading order in the large $N$ expansion, where planar graphs dominate,
we may interpret each term as a tessellation of the sphere by polygons in $q$ weighted colours,
corresponding to the $q$ different matrices. Given $\sigma\in S_q/(S_p\times S_{q-p})$, we define
an associated resolvent function
\begin{eqnarray}
    \label{discfct}
    W_{(p|\sigma)}(z)=\frac{1}{N}\left\langle\Tr\frac{1}{z-X_{(p|\sigma)}}\right\rangle\ ,\quad X_{(p|\sigma)}=\sum_{i=1}^pX_{\sigma(i)}\ ,
\end{eqnarray}
In the $S_q$-invariant case, for given $p$, all $|S_q/(S_p\times S_{q-p})|={q \choose p}$ partition
functions $W_{(p|\sigma)}(z)$ are described by the same function, so that we henceforth abbreviate
$W_{(p)}(z):=W_{(p|\sigma_0)}(z)$ for a representative $\sigma_0$. For $p=1$, our definition of
$W_{(p)}(z)$ reduces to the one studied in
\cite{Kazakov1988,Daul:1994qy,ZinnJustin:1999jg,Bonnet:1999nf,Eynard:1999gp,kulanthaivelu2019freevariableloopequations}. These functions
generate graphs with a single connected boundary containing $p$ equally weighted colours. The
matrix, and hence spin, used to define the resolvent defines the set of spins allowed on the
boundary. Therefore we term the resolvent of a single matrix the \textit{fixed} boundary condition.
Similarly we call the resolvent for the sum of all matrices the \textit{free} boundary condition,
and we call the resolvent for the sum of $p$ random matrices the \textit{partially magnetised} or
\textit{mixed} boundary condition.

Our goal in the following is to describe the planar resolvents for these boundary conditions. We
present the curves as $F_{(p)}(x,y)=0$, with the coefficient convention introduced above. The
asymptotic expansions are again given in appendix~\ref{app:asympt}.

\subsection{3-state Potts model}
\subsubsection{$p=1$}

The potential function is given by \beq \label{3Potts_p=1} V(\mathbf{X}) = \sum_{i=0}^2
    \left(\frac{t_2-1}{2}X_i^2 + \frac{t_3}{3}X_i^3\right) - X_0 X_1 -X_1 X_2 - X_2 X_0.
\eeq
The symmetry group for this model is $S_3\times\mathbb{Z}_2$, corresponding to permutation symmetry
of the matrices, and the discrete symmetry $X_i \rightarrow X_i^T \; \forall i$. With
$n_{\mathrm{max}} = 5$ the total number of inequivalent loop equations is 64. We have $\sum_{i=0}^5
    |\mathcal{W}_i| = 54$ resolvent functions. This is a closed and consistent system of polynomial
equations and so many equations are not independent. Therefore only a subset of these are required
to determine the resolvent. Taking $F_{(1)}\!\left(z,t_3z^2+t_2z-W_{(1)}(z)\right)=0$ we get

\begin{align}
    \label{curve:potts_fixed}
    F_{(1)}(x,y)={} & y^{5}{}+y^{4}\left(-t_3x^{2} + \bigl(-t_2-3\bigr)x + \frac{18 t_2-17}{4 t_3}\right)\notag \\
                    & {}+y^{3}\left(3 t_3x^{3} - \frac{3}{2} \left(t_2-5\right)x^{2} + \frac{-9 t_2^2-12 t_2+2 t_3^2+9}{2 t_3}x + c_{0,3}\right)\notag \\
                    & {}+y^{2}\left(-\frac{13 t_3}{4}x^{4} + \bigl(7 t_2-6\bigr)x^{3} - \frac{-15 t_2^2-54 t_2+10 t_3^2+9}{4 t_3}x^{2}\right.\notag \\*
                    & \qquad{}\left.{}- \frac{13 t_2^3-4 t_3^2 t_2-9 t_2+6 t_3^2}{2 t_3^2}x + c_{0,2}\right)\notag \\
                    & {}+y\left(\frac{3 t_3}{2}x^{5} - \frac{3}{2} \left(4 t_2-1\right)x^{4} + \frac{3 t_2^2-6 t_2+2 t_3^2}{t_3}x^{3}\right.\notag \\*
                    & \qquad{}\left.{}+ \frac{3 \left(5 t_2^3+3 t_2^2-3 t_3^2 t_2+t_3^2\right)}{2 t_3^2}x^{2} - \frac{6 t_2^4-6 t_2^3+6 t_3^2 t_2-t_3^4}{2 t_3^3}x + c_{0,1}\right)\notag \\
                    & {}-\frac{t_3}{4}x^{6} + \frac{3 t_2}{2}x^{5} - \frac{9 t_2^2+2 t_3^2}{4 t_3}x^{4} - \frac{t_2 \left(t_2^2-2 t_3^2\right)}{t_3^2}x^{3} + \frac{12 t_2^4-6 t_3^2 t_2^2-t_3^4}{4 t_3^3}x^{2}\notag \\
                    & \qquad{}- \frac{t_2 \left(2 t_2^2-t_3^2\right)}{2 t_3^2}x + c_{0,0}.
\end{align}

\subsubsection{$p=2$}
In order to compute the planar resolvent for the mixed boundary condition we change the basis of
the matrix integral \eref{3Potts_p=1} to $M_0 = X_1 + X_2, M_1 = X_2 + X_0, M_2 = X_0 + X_1$. Under
this change of basis the potential becomes
\begin{eqnarray}
    \label{mixedAction}
    V(\mathbf{M}) & = & \sum_{i=0}^2 \frac{3t_2 - 1}{8}  M_i^2  - \frac{t_2 + 1}{4} \left( M_0 M_1 + M_1 M_2 + M_2 M_0 \right) \\
                  &   & \quad + \, \frac{t_3}{8} \big( M_0^2 M_1 + M_0^2 M_2 + M_1^2 M_0 + M_1^2 M_2 \nonumber \\
                  &   & \quad\quad + \, M_2^2 M_0 + M_2^2 M_1 - 3 M_0 M_1 M_2 - 3 M_2 M_1 M_0 \big)
    +\frac{t_3}{24}\sum_{i=0}^2 M_i^3.  \nonumber
\end{eqnarray}

\noindent The resulting potential retains the $S_3\times\mathbb{Z}_2$ symmetry of \eref{3Potts_p=1}, but
possesses many more interaction terms. The choice of basis is arbitrary. In this case our
goal is to compute
\begin{equation}
    W_{(2)}(z) = \frac{1}{N} \left\langle \Tr\, \frac{1}{z-M_0} \right\rangle.
\end{equation}

In contrast to the fixed-spin boundary condition, we are able to determine a closed set of
algebraic constraints on the planar resolvent only if we extend to level $n_{\mathrm{max}} = 6$
loop equations. The increased number of terms in the potential relative to \eref{3Potts_p=1}
results in a greater number of resolvent functions present in the loop equations. At this level we
have $\sum_{i=0}^6|\mathcal{W}_i|=144$ resolvent functions and $\sum_{i=0}^6|\mathcal{V}_i|=245$
equations. Once redundancies have been removed, these loop equations contain a closed system of
equations. We have not determined a minimal closed subsystem.

The final expression for the spectral curve,     $F_{(2)}\!\left(z,W_{(2)}(z)\right)=0$, is degree $10$ in the resolvent $W_{(2)}(z)$ and degree
$13$ in the boundary cosmological constant $z$. 

\begin{samepage}
\begin{align}
    \label{curve:potts_mixed}
    F_{(2)}(x,y)={} & \frac{4}{t_3^2}y^{10}
    +y^9\left(-\frac{4}{t_3}x^2-\frac{8(t_2-10)}{t_3^2}x
    +\frac{8(7t_2+1)}{t_3^3}\right)\notag \\*
                    & {}+y^8\left(\frac{3}{2}x^4+\frac{6(t_2-12)}{t_3}x^3+\cdots\right)
    +y^7\left(-\frac{t_3}{4}x^6-\frac{3}{2}(t_2-16)x^5+\cdots\right)\notag \\*
                    & {}+y^6\left(\frac{t_3^2}{64}x^8+\frac{(t_2-28)t_3}{8}x^7+\cdots\right)
    +y^5\left(\frac{3t_3^2}{16}x^9+\frac{(27t_2-320)t_3}{16}x^8+\cdots\right)\notag \\*
                    & {}+y^4\left(\frac{7t_3^2}{8}x^{10}+\frac{5(7t_2-48)t_3}{4}x^9+\cdots\right)\notag \\*
                    & {}+y^3\left(2t_3^2x^{11}+2(11t_2-50)t_3x^{10}+\cdots\right)\notag \\*
                    & {}+y^2\left(\frac{9t_3^2}{4}x^{12}+(27t_2-88)t_3x^{11}+\cdots\right)
    +y\left(t_3^2x^{13}+(13t_2-32)t_3x^{12}+\cdots\right)\notag \\*
                    & {}-t_3^2x^{12}+\cdots.
\end{align}

\end{samepage}
\noindent Each ellipsis denotes lower powers of $x$ in the corresponding coefficient, including 
correlator-dependent terms. 

At first glance, this is rather surprising, as it
was shown in \cite{Niedner:2016skw} that the function $G_{(2)}^Y(z)$, which contains the
information of the eigenvalue density distribution, satisfies a spectral curve that is degree 5 in
$G_{(2)}^Y(z)$ and degree 6 in $z$. In Appendix B we reconcile these observations through detailed analysis of the analytic structure of
the two functions.

\subsubsection{$p=3$}

Let us choose a basis for the potential \eref{3Potts_p=1} given by $M_0 = X_0 + X_1 + X_2, M_1 =
    X_0 - X_1 + X_2, M_2 = X_0 + X_1- X_2$. Then computing the resolvent corresponding to the free
boundary condition, $W_{(3)}(z)$, is equivalent to computing the planar resolvent of the matrix
$M_0$ with the following potential,
\begin{eqnarray}
    \label{freeAction}
    V(\mathbf{M}) & = & \frac{(t_2-2)}{4}M_0^2 + \frac{t_2}{4}(M_1^2 + M_2^2) + \frac{t_2}{4}(- M_0 M_1 + M_1 M_2 - M_2 M_0) \nonumber \\
                  &   & + \, \frac{t_3}{8}(M_1^2 M_0 + M_2^2 M_0 + M_1^2 M_2 + M_2^2 M_1) \\
                  &   & + \, \frac{t_3}{8}(- M_0^2 M_1 - M_0^2 M_2) + \frac{t_3}{12} M_0^3. \nonumber
\end{eqnarray}
The symmetry group of \eref{freeAction} possesses a $\mathbb{Z}_2\times\mathbb{Z}_2$ symmetry, with
one $\mathbb{Z}_2$ symmetry relating $M_1$ and $M_2$. However, the correlators also possess a
further set of constraints, owing to their origin in the 3-state Potts model, that are not manifest
in \eref{freeAction}. This full inherited $S_3$ symmetry gives additional relations, including
\begin{equation}
    \label{extraCon}
    W_{1}(z) = \frac{1}{3} \Delta W_{(3)} (z),
\end{equation}
which is easily deduced by rewriting the terms in the original basis and imposing the full $S_3$
symmetry in the expectation value.

As in the case of the mixed boundary condition, one must extend to level 6 to find a complete set
of constraints. Applying the elimination procedure and \eref{extraCon}, one can determine the
spectral curve for the free spin resolvent, given by an algebraic equation that is degree 5 in $z$
and degree 5 in the resolvent $W_{(3)}(z)$. Furthermore, with the relation
\begin{equation}
    \label{strange}
    G_{(3)}^Y(z) = z + W_{(3)}(z),
\end{equation}
one recovers the spectral curve for $G_{(3)}^Y(z)$ of \cite{Atkin:2015ksy}, $
    F_{(3)}\!\left(z,z+W_{(3)}(z)\right)=0$, where

\begin{align}
    \label{curve:potts_free}
    F_{(3)}(x,y)={} & \frac{108}{t_3}y^{5}{}+y^{4}\left(-12x^{2} - \frac{36 \left(t_2-1\right)}{t_3}x + \frac{27 \left(26 t_2-9\right)}{t_3^2}\right)\notag \\
                    & {}+y^{3}\left(-4x^{3} - \frac{18 \left(5 t_2+3\right)}{t_3}x^{2} - \frac{6 \left(39 t_2^2+2 t_3^2-27\right)}{t_3^2}x + c_{0,3}\right)\notag \\
                    & {}+y^{2}\left(9x^{4} + \frac{36 \left(t_2-2\right)}{t_3}x^{3} + \frac{9 \left(-15 t_2^2-54 t_2+2 t_3^2+9\right)}{t_3^2}x^{2}\right.\notag \\*
                    & \qquad{}\left.{}- \frac{54 \left(9 t_2^3+12 t_2^2-9 t_2+2 t_3^2\right)}{t_3^3}x + c_{0,2}\right)\notag \\
                    & {}+y\left(6x^{5} + \frac{18 \left(4 t_2-1\right)}{t_3}x^{4} + \frac{12 \left(21 t_2^2-18 t_2+2 t_3^2\right)}{t_3^2}x^{3}\right.\notag \\*
                    & \qquad{}\left.{}+ \frac{54 \left(3 t_2^3-15 t_2^2+3 t_3^2 t_2-t_3^2\right)}{t_3^3}x^{2}\right.\notag \\*
                    & \qquad{}\left.{}- \frac{18 \left(18 t_2^4+54 t_2^3-12 t_3^2 t_2^2+18 t_3^2 t_2-t_3^4\right)}{t_3^4}x + c_{0,1}\right)\notag \\
                    & {}+x^{6} + \frac{18 t_2}{t_3}x^{5} + \frac{3 \left(39 t_2^2+2 t_3^2\right)}{t_3^2}x^{4} + \frac{36 t_2 \left(9 t_2^2+2 t_3^2\right)}{t_3^3}x^{3}\notag \\
                    & \qquad{}+ \frac{9 \left(36 t_2^4+30 t_3^2 t_2^2+t_3^4\right)}{t_3^4}x^{2} + \frac{54 t_2 \left(6 t_2^2+t_3^2\right)}{t_3^3}x + c_{0,0}.
\end{align}

\subsection{General 3-state Potts}
We finally deploy our method to the 3-state Potts model with unequal cubic potentials.
\beq
    \label{3PottsGeneral_p=1}
    \begin{aligned}
        V(\mathbf{X}) ={} & \frac{t_2-1}{2}X_0^2 + \frac{s_2-1}{2}X_1^2 + \frac{r_2-1}{2}X_2^2 \\
                          & {}+ \frac{t_3}{3}X_0^3 + \frac{s_3}{3}X_1^3 + \frac{r_3}{3}X_2^3
        - X_0 X_1 - X_1 X_2 - X_2 X_0
    \end{aligned}
\eeq
In this case we have broken the $S_3$ symmetry. A slightly simpler variant of this was considered in \cite{Atkin:2012fx} but to our knowledge no complete relation has been written down. Like in the symmetric $p=1$ case we show $F_{(1)}\!\left(z,t_3z^2+t_2z-W_{(1)}(z)\right)=0$.
\begin{samepage}
\begin{align}
    \label{curve:potts_general}
    F_{(1)}(x,y)={} & y^7
    +y^6\left(-t_3x^2-(t_2+5)x+\frac{2t_2-1}{4t_3}
    +\frac{3r_2-2}{r_3}+\frac{3s_2-2}{s_3}\right)\notag \\*
                    & {}+y^5\left(5t_3x^3+\biggl(\frac{9t_2+21}{2}-\frac{(3r_2-2)t_3}{r_3}
    -\frac{(3s_2-2)t_3}{s_3}\biggr)x^2+\cdots\right)\notag \\*
                    & {}+y^4\left(-\frac{41t_3}{4}x^4-\biggl(8t_2+12-\frac{(25r_2-12)t_3}{2r_3}
    -\frac{(25s_2-12)t_3}{2s_3}\biggr)x^3+\cdots\right)\notag \\*
                    & {}+y^3\left(11t_3x^5+\biggl(7t_2+8-\frac{(41r_2-13)t_3}{2r_3}
    -\frac{(41s_2-13)t_3}{2s_3}\biggr)x^4+\cdots\right)\notag \\*
                    & {}+y^2\left(-\frac{13t_3}{2}x^6-\biggl(3t_2+3-\frac{3(11r_2-2)t_3}{2r_3}
    -\frac{3(11s_2-2)t_3}{2s_3}\biggr)x^5+\cdots\right)\notag \\*
                    & {}+y\left(2t_3x^7+\biggl(\frac{t_2+1}{2}-\frac{(13r_2-1)t_3}{2r_3}
    -\frac{(13s_2-1)t_3}{2s_3}\biggr)x^6+\cdots\right)\notag \\*
                    & {}-\frac{t_3}{4}x^8
    +t_3\left(\frac{r_2}{r_3}+\frac{s_2}{s_3}\right)x^7+\cdots.
\end{align}

\end{samepage}
\noindent Unlike the symmetric case we find the degree of the curve to be $(\deg_xF_{(1)},\deg_yF_{(1)})=(8,7)$. However one still finds that in the symmetric limit $r_2=s_2=t_2$ and $r_3=s_3=t_3$, the physical branch reproduces the fixed-boundary resolvent of Section 6.1.1.

\section{Discussion}

In this work we developed an algorithm for systematically computing and solving the loop equations
of general multi-matrix models, with the goal of finding the spectral curve for the planar
resolvent of a given matrix degree of freedom. We demonstrated the utility of the approach by deploying
it to solve several unsolved matrix models described in the literature. We then focused our
attention on the 3-state Potts model, where we successfully computed the planar resolvents
describing fixed, mixed, and free boundary conditions using this technique. While the fixed and
free boundary conditions were manifestly in agreement with the literature, more care was required
for the mixed boundary condition. However, through studying the analytic structure predicted by the
saddle-point solution this too is consistent with the results of Atkin et al. We finally relaxed
the $S_3$ symmetry of the 3-state Potts model and demonstrated that this is solvable.

There are many ways to continue this work. Our procedure is algorithmic in nature, and so it would
be interesting to apply this to a wide variety of models for which the planar resolvent is known or
unknown. Determining a general principle that governs the solvability of correlators in the
multi-matrix models would be an exciting challenge. It would also be useful to connect this
approach with the matrix bootstrap
\cite{Lin_2020,khalkhali2025bootstrappingcriticalbehaviormultimatrix}: the algebraic reduction
developed here could reduce the number of independent moments entering its positivity constraints.
A further direction is to apply this framework to the matrix-model realisations of the $ADE$
minimal strings \cite{Rodriguez:2025ade} and investigate their planar spectral curves and boundary
observables. Work is in progress to develop a general-purpose software library for
generating and analysing loop equations, with the aim of making the framework applicable to
arbitrary matrix models with polynomial potentials.

An immediate offshoot of this work would be the calculation of planar resolvents describing boundary
conditions that interpolate between the fixed, mixed, and free boundary states in the 3-state Potts model, analogous to
\cite{Carroll:1997tr,Atkin:2012fx}. This would allow us to directly study the boundary
renormalisation group flow relating the different boundary states, which is expected to induce a
partial ordering of the spectrum of boundary states, conforming with the boundary analogue of the
$c$-theorem \cite{Zamolodchikov:1986gt}, which was first conjectured in \cite{Affleck:1991tk}, and
subsequently proven by Friedan and Konechny \cite{Friedan:2003yc}. It would also be interesting to
see whether one could compute the New boundary condition, which is likely given by a non-standard
resolvent function \cite{Kulanthaivelu:2019atg}, using this framework.

Finally, it would be fruitful to extend our work to the calculation of observables in multi-matrix
models beyond the planar limit. It is well known that the planar resolvent forms the initial data
for higher-genus and multi-boundary correlators in the one- and two-matrix models
\cite{Eynard:2002kg,Eynard:2004mh,Eynard:2008we}, and that this may be extended to certain
multi-matrix models \cite{Borot:2009ia}. It would be interesting to see whether one could reframe
the method of loop equations for various boundary conditions in the 3-state Potts model such that
one can calculate higher-order correlation functions in an analogous topological recursion
procedure. This would allow, for example, the computation of cylinder amplitudes for different
boundary states.

\acknowledgments
The majority of this work was written during my DPhil at the University of
Oxford, supported by the STFC grant ST/N504233/1. Substantial parts of the results and exposition presented here appeared previously in my DPhil thesis \cite{kulanthaivelu2020spin}. I thank John F.\ Wheater for his guidance,
many helpful discussions, and continued encouragement.


\clearpage
\appendix
\section{Analytic Structure and Asymptotics}
\label{app:asympt}

In this appendix we list the asymptotic expansions sheet by sheet, with coefficients in the
parameters of the corresponding potential, retaining terms through $z^{-1}$. The physical sheet is
labelled $0$ and has $W(z)_0=z^{-1}+\mathcal O(z^{-2})$. The remaining labels enumerate the local
branches at infinity, keeping ramified families together. The labels here do not prescribe the finite branch cuts. The
accompanying notebooks contain the complete calculations \cite{Kulanthaivelu:LoopEquationsCode}.

\subsection{Generic Cubic Two-Matrix Model}
The four sheets of the generic cubic two-matrix curve have the expansions
\begin{sheetasymptotics}
    \begin{align*}
        \sheetlabel{W(z)_{0}} & ={}z^{-1}+\mathcal O(z^{-2}) \\[4pt]
        \sheetlabel{W(z)_{1}} & ={}\frac{2 \alpha \left(- d^{2} + b e\right) + 2 \beta \left(- b^{2} + a d\right) + 2 \gamma \left(a e - b d\right)}{\left(- d^{2} + b e\right) \left(- \gamma^{2} + \alpha \beta\right)}z \\*
                              & \quad{}+\frac{\left(\beta d + e \gamma\right) \left(\alpha \left(- d^{2} + b e\right) + \beta \left(- b^{2} + a d\right) + \gamma \left(a e - b d\right)\right)}{\left(- d^{2} + b e\right)^{2} \left(- \gamma^{2} + \alpha \beta\right)^{2}}-z^{-1}+\mathcal O(z^{-2}) \\[4pt]
        \sheetlabel{W(z)_{2}} & ={}\frac{- 2 d^{3} - 2 \left(d^{2} - b e\right)^{3/2} - a e^{2} + 3 b d e}{e^{2}}z^{2}+\frac{\alpha e^{2} + \beta \left(2 d^{2} - b e\right) - \sqrt{d^{2} - b e} \left(- 2 \beta d - 2 e \gamma\right) + 2 d e \gamma}{e^{2} \left(- \gamma^{2} + \alpha \beta\right)}z \\*
                              & \quad{}-\frac{\left(\beta d + \beta \sqrt{d^{2} - b e} + e \gamma\right)^{2}}{4 e^{2} \sqrt{d^{2} - b e} \left(- \gamma^{2} + \alpha \beta\right)^{2}}+\mathcal O(z^{-2}) \\[4pt]
        \sheetlabel{W(z)_{3}} & ={}\frac{- 2 d^{3} + 2 \left(d^{2} - b e\right)^{3/2} - a e^{2} + 3 b d e}{e^{2}}z^{2}+\frac{\alpha e^{2} + \beta \left(2 d^{2} - b e\right) - \sqrt{d^{2} - b e} \left(2 \beta d + 2 e \gamma\right) + 2 d e \gamma}{e^{2} \left(- \gamma^{2} + \alpha \beta\right)}z \\*
                              & \quad{}+\frac{\left(\beta \sqrt{d^{2} - b e} - \beta d - e \gamma\right)^{2}}{4 e^{2} \sqrt{d^{2} - b e} \left(- \gamma^{2} + \alpha \beta\right)^{2}}+\mathcal O(z^{-2})
    \end{align*}
\end{sheetasymptotics}

\subsection{Quartic and cubic multi-matrix models}

\subsubsection{Quartic two-matrix models}
For $(\alpha,\beta,\gamma)=(g,g,g)$ and $(g,-g,g)$, the four sheets are
\begin{sheetasymptotics}
    \begin{align*}
        \sheetlabel{W_A(z)_{0}} & ={}z^{-1}+\mathcal O(z^{-2}) \\[4pt]
        \sheetlabel{W_A(z)_{1}} & ={}-z^{-1}+\mathcal O(z^{-2}) \\[4pt]
        \sheetlabel{W_A(z)_{2}} & ={}-z^{-1}+\mathcal O(z^{-2}) \\[4pt]
        \sheetlabel{W_A(z)_{3}} & ={}gz^{3}+z+z^{-1}+\mathcal O(z^{-2})
    \end{align*}
\end{sheetasymptotics}

\Needspace{7\baselineskip}
\noindent For $(\alpha,\beta,\gamma)=(g,g,-g)$, the six sheets are
\begin{sheetasymptotics}
    \begin{align*}
        \sheetlabel{W_A(z)_{0}} & ={}z^{-1}+\mathcal O(z^{-2}) \\[4pt]
        \sheetlabel{W_A(z)_{1}} & ={}gz^{3}+z+z^{-1}+\mathcal O(z^{-2}) \\[4pt]
        \sheetlabel{W_A(z)_{2}} & ={}3z-z^{-1}+\mathcal O(z^{-2}) \\[4pt]
        \sheetlabel{W_A(z)_{3}} & ={}4z+\frac{2 \sqrt[3]{2}}{\sqrt[3]{g}}z^{1/3}+\frac{2^{2/3}}{g^{2/3}}z^{-1/3}-\frac{1}{3}z^{-1}+\mathcal O(z^{-5/3}) \\[4pt]
        \sheetlabel{W_A(z)_{4}} & ={}4z+\frac{\sqrt[3]{2} \left(-1 - i \sqrt{3}\right)}{\sqrt[3]{g}}z^{1/3}+\frac{2^{2/3} \left(- \frac{1}{2} + \frac{i \sqrt{3}}{2}\right)}{g^{2/3}}z^{-1/3}-\frac{1}{3}z^{-1}+\mathcal O(z^{-5/3}) \\[4pt]
        \sheetlabel{W_A(z)_{5}} & ={}4z+\frac{\sqrt[3]{2} \left(-1 + i \sqrt{3}\right)}{\sqrt[3]{g}}z^{1/3}+\frac{2^{2/3} \left(- \frac{1}{2} - \frac{i \sqrt{3}}{2}\right)}{g^{2/3}}z^{-1/3}-\frac{1}{3}z^{-1}+\mathcal O(z^{-5/3})
    \end{align*}
\end{sheetasymptotics}

\Needspace{7\baselineskip}
\noindent For $(\alpha,\beta,\gamma)=(-g,g,g)$, the six sheets are the same as in the $(g,g,-g)$ case but with $g\rightarrow -g$.

\subsubsection{Cubic three-matrix model}
\begin{sheetasymptotics}
    \begin{align*}
        \sheetlabel{W_{A_0}(z)_{0}} & ={}z^{-1}+\mathcal O(z^{-2}) \\[4pt]
        \sheetlabel{W_{A_0}(z)_{1}} & ={}-z^{-1}+\mathcal O(z^{-2}) \\[4pt]
        \sheetlabel{W_{A_0}(z)_{2}} & ={}-z^{-1}+\mathcal O(z^{-2}) \\[4pt]
        \sheetlabel{W_{A_0}(z)_{3}} & ={}9 gz^{2}+9z+\frac{2}{g}+z^{-1}+\mathcal O(z^{-2}) \\[4pt]
        \sheetlabel{W_{A_0}(z)_{4}} & ={}gz^{2}+z+z^{-1}+\mathcal O(z^{-2}) \\[4pt]
        \sheetlabel{W_{A_0}(z)_{5}} & ={}6z-\frac{2 \sqrt{6}}{\sqrt{g}}z^{1/2}+\frac{1}{g}-\frac{\sqrt{6}}{24 g^{3/2}}z^{-1/2}-\frac{1}{2}z^{-1}+\mathcal O(z^{-3/2}) \\[4pt]
        \sheetlabel{W_{A_0}(z)_{6}} & ={}6z+\frac{2 \sqrt{6}}{\sqrt{g}}z^{1/2}+\frac{1}{g}+\frac{\sqrt{6}}{24 g^{3/2}}z^{-1/2}-\frac{1}{2}z^{-1}+\mathcal O(z^{-3/2})
    \end{align*}
\end{sheetasymptotics}

\Needspace{10\baselineskip}
\subsection{3-state Potts Model}

\subsubsection{Fixed boundary conditions}

The asymptotic behaviour of the planar resolvent $W_{(1)}(z)$ is given by:
\begin{sheetasymptotics}
    \begin{align*}
        \sheetlabel{W_{(1)}(z)_0} & ={}   \frac{1}{z} + \mathcal{O}(z^{-2}) \\[4pt]
        \sheetlabel{W_{(1)}(z)_1} & ={}  t_3 z^2 + (t_2 -1)z - 2t_3^{-1/2} z^{1/2} + \frac{t_2 -2}{t_3} - \frac{(t_2-2)^2}{4t_3^{3/2}} z^{-1/2} + \mathcal{O}(z^{-3/2}) \\[4pt]
        \sheetlabel{W_{(1)}(z)_2} & ={}  t_3 z^2 + (t_2 -1)z + 2t_3^{-1/2}z^{1/2} + \frac{t_2 -2}{t_3} + \frac{(t_2-2)^2}{4t_3^{3/2}} z^{-1/2} + \mathcal{O}(z^{-3/2}) \\[4pt]
        \sheetlabel{W_{(1)}(z)_3} & ={}  t_3 z^2 + \left(t_2-\frac12\right) z - \frac{t_3^{-1/2}}{2\sqrt{2}} z^{1/2} + \frac{(10t_2-1)}{8t_3} \\*
                                  & \quad{}- \frac{(4t_2^2 - 20t_2 +1)} {32\sqrt{2}t_3^{3/2}}z^{-1/2} - \frac{1}{2} z^{-1} + \mathcal{O}(z^{-3/2}) \\[4pt]
        \sheetlabel{W_{(1)}(z)_4} & ={}  t_3 z^2 + \left(t_2-\frac12\right) z + \frac{t_3^{-1/2}}{2\sqrt{2}} z^{1/2} + \frac{(10t_2-1)}{8t_3} \\*
                                  & \quad{}+ \frac{(4t_2^2 - 20t_2 +1)}{32\sqrt{2}t_3^{3/2}}z^{-1/2} - \frac{1}{2} z^{-1} + \mathcal{O}(z^{-3/2})
    \end{align*}
\end{sheetasymptotics}%

\subsubsection{Mixed boundary conditions}
\label{app:mixed-asymptotics}

The asymptotic behaviour of the planar resolvent $W_{(2)}(z)$ is given by:
\begin{sheetasymptotics}
    \begin{align*}
        \sheetlabel{W_{(2)}(z)_0} & ={}   z^{-1} + \mathcal{O}(z^{-2}) \\[4pt]
        \sheetlabel{W_{(2)}(z)_1} & ={} \frac{t_3 z^2}{4}+\frac{\left(t_2 - 2\right)}{2}z -t_3^{-1/2}z^{1/2} +\frac{t_2-1}{2 t_3}-\frac{(t_2-1)^2}{8 t_3^{3/2}}z^{-1/2}+\frac{1}{2}z^{-1} + \mathcal{O}(z^{-3/2}) \\[4pt]
        \sheetlabel{W_{(2)}(z)_2} & ={} \frac{t_3 z^2}{4}+\frac{\left(t_2 - 2\right)}{2}z+t_3^{-1/2}z^{1/2}+\frac{t_2-1}{2 t_3}+\frac{(t_2-1)^2}{8 t_3^{3/2}}z^{-1/2}+\frac{1}{2}z^{-1}+ \mathcal{O}(z^{-3/2}) \\[4pt]
        \sheetlabel{W_{(2)}(z)_3} & ={} -2 z + (1+i)t_3^{-1/2}z^{1/2} -\frac{2 t_2}{t_3}  +\frac{\left(1-i\right) \left(t_2^2-(6-4 i) t_2+1\right)}{8t_3^{3/2}}z^{-1/2} + \mathcal{O}(z^{-3/2}) \\[4pt]
        \sheetlabel{W_{(2)}(z)_4} & ={} -2 z -(1-i)t_3^{-1/2}z^{1/2} -\frac{2 t_2}{t_3}  - \frac{\left(1+i\right) \left(t_2^2-(6+4 i) t_2+1\right)}{8t_3^{3/2}}z^{-1/2} + \mathcal{O}(z^{-3/2}) \\[4pt]
        \sheetlabel{W_{(2)}(z)_5} & ={} -2 z -(1+i)t_3^{-1/2}z^{1/2} -\frac{2 t_2}{t_3}  -\frac{\left(1-i\right) \left(t_2^2-(6-4 i) t_2+1\right)}{8t_3^{3/2}}z^{-1/2}+ \mathcal{O}(z^{-3/2}) \\[4pt]
        \sheetlabel{W_{(2)}(z)_6} & ={} -2 z + (1-i) t_3^{-1/2}z^{1/2} -\frac{2 t_2}{t_3}  +\frac{\left(1+i\right) \left(t_2^2-(6+4 i) t_2+1\right)}{8t_3^{3/2}}z^{-1/2} + \mathcal{O}(z^{-3/2}) \\[4pt]
        \sheetlabel{W_{(2)}(z)_7} & ={} \frac{t_3 z^2}{4}+\frac{1}{2} \left(t_2-6\right) z+it_3^{-1/2}z^{1/2}-\frac{3 t_2+1}{2 t_3}-\frac{i \left(t_2^2-10 t_2+1\right)}{8 t_3^{3/2}}z^{-1/2} \\*
                                  & \quad{}-\frac{1}{2}z^{-1} + \mathcal{O}(z^{-3/2}) \\[4pt]
        \sheetlabel{W_{(2)}(z)_8} & ={} \frac{t_3 z^2}{4}+\frac{1}{2} \left(t_2-6\right) z-it_3^{-1/2}z^{1/2}-\frac{3 t_2+1}{2 t_3} +\frac{i \left(t_2^2-10 t_2+1\right)}{8 t_3^{3/2}}z^{-1/2} \\*
                                  & \quad{}-\frac{1}{2}z^{-1}  +\mathcal{O}(z^{-3/2}) \\[4pt]
        \sheetlabel{W_{(2)}(z)_9} & ={} -4z - \frac{4t_{2}}{t_3} - z^{-1} + \mathcal{O}(z^{-2})
    \end{align*}
\end{sheetasymptotics}%

\subsubsection{Free boundary conditions}

The asymptotic behaviour of the planar resolvent $W_{(3)}(z)$ is given by:
\begin{sheetasymptotics}
    \begin{align*}
        \sheetlabel{W_{(3)}(z)_0} & ={} z^{-1} + \mathcal{O}(z^{-2}) \\[4pt]
        \sheetlabel{W_{(3)}(z)_1} & ={} \frac{t_3}{9} z^2 + \frac{(t_2-3)}{3} z + \frac{1}{z} + \mathcal{O}(z^{-2}) \\[4pt]
        \sheetlabel{W_{(3)}(z)_2} & ={} -\frac{3}{2} z + \frac{3i}{2\sqrt{2}t_3^{1/2}}z^{1/2} - \frac{(18t_2 -9)}{8t_3} - \frac{3i(4t_2^2 - 36t_2 + 9)}{32\sqrt{2} t_3^{3/2}} z^{-1/2} \\*
                                  & \quad{}- \frac{1}{2} z^{-1} + \mathcal{O}(z^{-3/2}) \\[4pt]
        \sheetlabel{W_{(3)}(z)_3} & ={} -\frac{3}{2} z - \frac{3i}{2\sqrt{2}t_3^{1/2}}z^{1/2} - \frac{(18t_2 -9)}{8t_3} + \frac{3i(4t_2^2 - 36t_2 + 9)}{32\sqrt{2} t_3^{3/2}} z^{-1/2} \\*
                                  & \quad{}- \frac{1}{2} z^{-1} + \mathcal{O}(z^{-3/2}) \\[4pt]
        \sheetlabel{W_{(3)}(z)_4} & ={} - \frac{4}{3} z - \frac{2t_2}{t_3} - z^{-1} + \mathcal{O}(z^{-2})
    \end{align*}
\end{sheetasymptotics}%

\subsection{General 3-state Potts}
\begin{sheetasymptotics}
    \begin{align*}
        \sheetlabel{W_{(1)}(z)_{0}} & ={}z^{-1}+\mathcal O(z^{-2}) \\[4pt]
        \sheetlabel{W_{(1)}(z)_{1}} & ={}t_3z^{2}+\left(-1 + t_2\right)z-\left(\frac{1}{\sqrt{r_3}} + \frac{1}{\sqrt{s_3}}\right)z^{1/2}+\frac{-1 + r_2}{2 r_3} + \frac{-1 + s_2}{2 s_3} - \frac{1}{\sqrt{r_3} \sqrt{s_3}} \\*
                                    & \quad{}-\left(\frac{\left(-1 + r_2\right)^{2}}{8 r_3^{3/2}} + \frac{\left(-1 + s_2\right)^{2}}{8 s_3^{3/2}} + \frac{3 - 2 r_2}{8 r_3 \sqrt{s_3}} + \frac{3 - 2 s_2}{8 \sqrt{r_3} s_3}\right)z^{-1/2}+\mathcal O(z^{-3/2}) \\[4pt]
        \sheetlabel{W_{(1)}(z)_{2}} & ={}t_3z^{2}+\left(-1 + t_2\right)z+\left(\frac{1}{\sqrt{r_3}} + \frac{1}{\sqrt{s_3}}\right)z^{1/2}+\frac{-1 + r_2}{2 r_3} + \frac{-1 + s_2}{2 s_3} - \frac{1}{\sqrt{r_3} \sqrt{s_3}} \\*
                                    & \quad{}+\left(\frac{\left(-1 + r_2\right)^{2}}{8 r_3^{3/2}} + \frac{\left(-1 + s_2\right)^{2}}{8 s_3^{3/2}} + \frac{3 - 2 r_2}{8 r_3 \sqrt{s_3}} + \frac{3 - 2 s_2}{8 \sqrt{r_3} s_3}\right)z^{-1/2}+\mathcal O(z^{-3/2}) \\[4pt]
        \sheetlabel{W_{(1)}(z)_{3}} & ={}t_3z^{2}+\left(-1 + t_2\right)z-\left(\frac{1}{\sqrt{s_3}} - \frac{1}{\sqrt{r_3}}\right)z^{1/2}+\frac{1}{\sqrt{r_3} \sqrt{s_3}} + \frac{-1 + r_2}{2 r_3} + \frac{-1 + s_2}{2 s_3} \\*
                                    & \quad{}-\left(- \frac{\left(-1 + r_2\right)^{2}}{8 r_3^{3/2}} + \frac{\left(-1 + s_2\right)^{2}}{8 s_3^{3/2}} - \frac{3 - 2 s_2}{8 \sqrt{r_3} s_3} + \frac{3 - 2 r_2}{8 r_3 \sqrt{s_3}}\right)z^{-1/2}+\mathcal O(z^{-3/2}) \\[4pt]
        \sheetlabel{W_{(1)}(z)_{4}} & ={}t_3z^{2}+\left(-1 + t_2\right)z+\left(\frac{1}{\sqrt{s_3}} - \frac{1}{\sqrt{r_3}}\right)z^{1/2}+\frac{1}{\sqrt{r_3} \sqrt{s_3}} + \frac{-1 + r_2}{2 r_3} + \frac{-1 + s_2}{2 s_3} \\*
                                    & \quad{}+\left(- \frac{\left(-1 + r_2\right)^{2}}{8 r_3^{3/2}} + \frac{\left(-1 + s_2\right)^{2}}{8 s_3^{3/2}} - \frac{3 - 2 s_2}{8 \sqrt{r_3} s_3} + \frac{3 - 2 r_2}{8 r_3 \sqrt{s_3}}\right)z^{-1/2}+\mathcal O(z^{-3/2}) \\[4pt]
        \sheetlabel{W_{(1)}(z)_{5}} & ={}t_3z^{2}+\left(- \frac{1}{2} + t_2\right)z+\frac{\sqrt{2}}{4 \sqrt{t_3}}z^{1/2}+\frac{r_2}{2 r_3} + \frac{s_2}{2 s_3} + \frac{-1 + 2 t_2}{8 t_3} \\*
                                    & \quad{}+\frac{\sqrt{2} \left(\left(-1 + 2 t_2\right)^{2} - 8 t_3 \left(\frac{r_2}{r_3} + \frac{s_2}{s_3}\right)\right)}{64 t_3^{3/2}}z^{-1/2}-\frac{1}{2}z^{-1}+\mathcal O(z^{-3/2}) \\[4pt]
        \sheetlabel{W_{(1)}(z)_{6}} & ={}t_3z^{2}+\left(- \frac{1}{2} + t_2\right)z-\frac{\sqrt{2}}{4 \sqrt{t_3}}z^{1/2}+\frac{r_2}{2 r_3} + \frac{s_2}{2 s_3} + \frac{-1 + 2 t_2}{8 t_3} \\*
                                    & \quad{}+\frac{\sqrt{2} \left(- \left(-1 + 2 t_2\right)^{2} + 8 t_3 \left(\frac{r_2}{r_3} + \frac{s_2}{s_3}\right)\right)}{64 t_3^{3/2}}z^{-1/2}-\frac{1}{2}z^{-1}+\mathcal O(z^{-3/2})
    \end{align*}
\end{sheetasymptotics}

\section{Mixed boundary saddle-point solution}
\label{app:mixed-saddle}

Here we show the degree-ten curve for $W_{(2)}$ is compatible with the degree-five curve for the
auxiliary function $G_{(2)}^Y$ obtained in \cite{Niedner:2016skw}. We make this comparison by
reducing the matrix integral to an eigenvalue problem and following the analytic continuation of
the mixed resolvent.

Set $X_\pm=X_1\pm X_2$, keeping $X_0$ as the third matrix in \eqref{3Potts_p=1}, and define
\begin{equation}
    V_+(x)=\frac{t_2-2}{4}x^2+\frac{t_3}{12}x^3,
    \qquad V_0(x)=\frac{t_2-1}{2}x^2+\frac{t_3}{3}x^3.
\end{equation}
The $X_-$ integral is Gaussian and can be integrated out. Diagonalising $X_+$ and $X_0$, with
eigenvalues $\lambda_i$ and $\mu_i$, and performing the angular integral using the HCIZ formula
\cite{Harish-Chandra538}, we obtain
\begin{equation}
    \label{MixedSaddle}
    Z\propto\int\prod_{i=1}^N d\lambda_i\,d\mu_i\,
    \frac{\Delta(\lambda)\Delta(\mu)\det_{i,j}e^{N\lambda_i\mu_j}}
    {\sqrt{\prod_i b_i}\prod_{i<j}(b_i+b_j)}
    e^{-N\sum_i\left(V_+(\lambda_i)+V_0(\mu_i)\right)},
\end{equation}
where $b_i=t_2+t_3\lambda_i$ and $\Delta$ is the Vandermonde determinant. Overall factors
independent of the eigenvalues have been suppressed.

Following \cite{Atkin:2015ksy}, write
\begin{equation}
    \label{mixed:G-definition}
    G_A^B(z)=\lim_{N\to\infty}\frac1N\frac{\partial}{\partial z}
    \log\left\langle\det_{i,j}e^{N\xi_i\eta_j}\right\rangle_{\xi_N=z},
\end{equation}
where $\xi_i$ and $\eta_j$ are the eigenvalues of $A$ and $B$, respectively. Define the reflection
$s(z)=-z-2t_2/t_3$. Stationarity of \eqref{MixedSaddle} in the planar limit gives
\begin{align}
    \label{2MMsaddle2}
    V_+'(z) & =\overline W_{(2)}(z)+\overline G_{X_+}^{X_0}(z)+W_{(2)}(s(z)), \\
    V_0'(z) & =\overline W_{(1)}(z)+\overline G_{X_0}^{X_+}(z),
\end{align}
on the corresponding eigenvalue cuts, with $\overline f(x)=\tfrac12\left(f(x+i0)+f(x-i0)\right)$
denoting the mean boundary value. The reflected resolvent comes from differentiating the Gaussian
determinant.

Let $G(z)=G_{(2)}^Y(z)$, where $Y$ is the auxiliary Gaussian matrix coupled to $X_0+X_1+X_2$ in
\cite{Atkin:2015ksy}. Its curve satisfies
\begin{equation}
    \mathcal F_{(2)}(z,G(z))=F_{(1)}(G(z)-z,G(z))=0.
\end{equation}
We use $\mathcal F_{(2)}$ for this auxiliary curve to distinguish it from the degree-ten resolvent
curve $F_{(2)}$. Denote the branch adjacent to the physical branch of $G$ by $G_0$. Combining the
saddle-point equations with the HCIZ branch relations gives
\begin{equation}
    \label{mixed:physical-relation}
    W_0(z)+W_0(s(z))+G_0(z)=U(z),
    \qquad U(z)=\frac{t_3}{4}z^2+\frac{t_2}{2}z,
\end{equation}
where $W_j(z)=W_{(2)}(z)_j$ and $W_0(z)\sim z^{-1}$. In particular, $U(s(z))=U(z)$ and
$G_0(s(z))=G_0(z)$.

\begin{figure}[htbp]
    \centering    \includegraphics[scale=0.85]{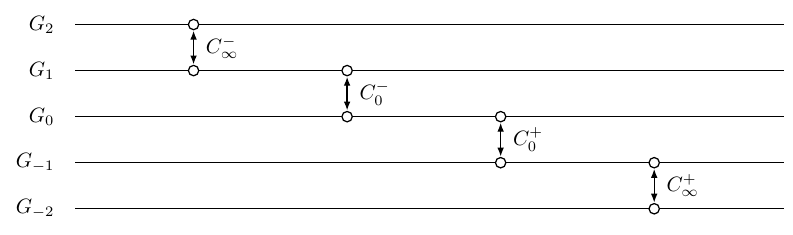}
    \caption{The five sheets of the auxiliary function $G$. Arrows join the sheets exchanged
        across each labelled cut; the reflection $s$ interchanges the $+$ and $-$ cuts.}
    \label{fig:mixed-G-continuation}
\end{figure}

In the one-cut phase, take $C_0^+=[a,b]$ and $C_\infty^+=[c,\infty)$, with $a<b<c$, and define
their reflected cuts $C_0^-=s(C_0^+)$ and $C_\infty^-=s(C_\infty^+)$. We assume these cuts are
disjoint. Continuing $G_0$ through $C_0^+$ gives $G_{-1}$, and then through $C_\infty^+$ gives
$G_{-2}$; the other two auxiliary sheets are $G_1(z)=G_{-1}(s(z))$ and $G_2(z)=G_{-2}(s(z))$.
Choose the square-root branches so that $G_{-1}(z)=z+t_3^{-1/2}z^{1/2}+O(1)$ and
$G_{-2}(z)=z-t_3^{-1/2}z^{1/2}+O(1)$, with $s(z)^{1/2}\sim iz^{1/2}$. The auxiliary sheet structure
is shown in Figure~\ref{fig:mixed-G-continuation}.

\begin{figure}[htbp]
    \centering
    \includegraphics[scale=0.85]{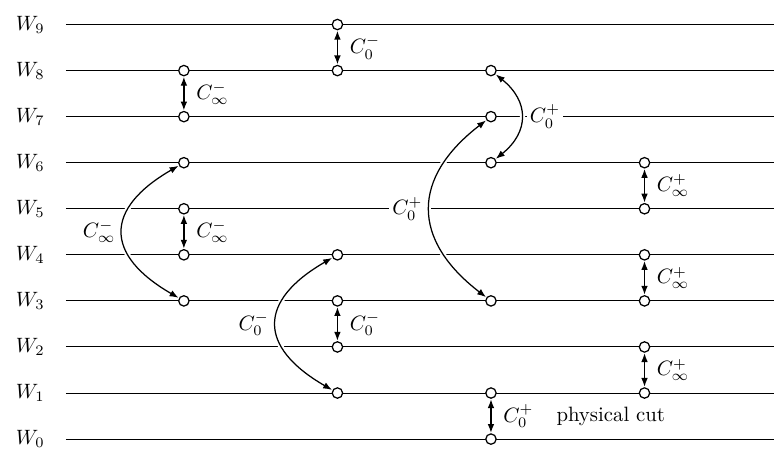}
    \caption{The ten sheets of $W_{(2)}$, with $W_0$ the physical sheet. Arrows with the same cut
        label act simultaneously. Only circled endpoints indicate sheet exchanges; crossings of
        intervening sheet lines do not.}
    \label{fig:mixed-W-continuation}
\end{figure}

Crossing $C_0^+$ in \eqref{mixed:physical-relation} changes $W_0$ to $W_1$ and $G_0$ to $G_{-1}$,
while $W_0(s(z))$ remains analytic. Repeating this continuation gives
\begin{equation}
    \label{mixed:sheet-relations}
    W_j(z)=U(z)-G_{\alpha_j}(z)-W_{\beta_j}(s(z)),
    \qquad
    \begin{array}{c|rrrrrrrrrr}
        j        & 0 & 1  & 2  & 3  & 4  & 5  & 6  & 7  & 8  & 9 \\
        \hline
        \alpha_j & 0 & -1 & -2 & -2 & -1 & -1 & -2 & -2 & -2 & -2 \\
        \beta_j  & 0 & 0  & 0  & 1  & 1  & 2  & 2  & 4  & 3  & 7
    \end{array}
\end{equation}
The sheet labels agree with the asymptotic catalogue in section~\ref{app:mixed-asymptotics} and
expanding \eqref{mixed:sheet-relations} at infinity reproduces the ten series. The complete
continuation is shown in Figure~\ref{fig:mixed-W-continuation}.
These exchanges close on ten resolvent sheets, consistent with our results.

\FloatBarrier
\section{Planar solutions and critical behaviour}
\label{sec:qtm-solutions}

In this section we show that the four quartic two-matrix models are more tractable than their
original interactions suggest. We show that their spectral curves can be grouped into two pairs. One pair reduces to the
ordinary quartic one-matrix problem, while the other admits the usual rational parameterisation \cite{Eynard:2002kg}. Routine methods then determine the remaining moments, the
critical couplings, and the string susceptibility.

We use the normalisation and notation of \cite{khalkhali2025bootstrappingcriticalbehaviormultimatrix} and label
the four sets of couplings by
\begin{equation}
    \mathrm I:(g,g,g),\qquad
    \mathrm{II}:(g,-g,g),\qquad
    \mathrm{III}:(g,g,-g),\qquad
    \mathrm{IV}:(-g,g,g).
    \label{eq:qtm-models}
\end{equation}
Write
\begin{equation}
    m_k^{\mathrm M}(g)
    =\lim_{N\to\infty}\frac1N
    \bigl\langle\operatorname{Tr}A^k\bigr\rangle_{\mathrm M}
    =\lim_{N\to\infty}\frac1N
    \bigl\langle\operatorname{Tr}B^k\bigr\rangle_{\mathrm M},
    \qquad \mathrm M\in\{\mathrm I,\mathrm{II},\mathrm{III},\mathrm{IV}\}.
    \label{eq:qtm-moment-definition}
\end{equation}
\noindent From the spectral curves we have
\begin{equation}
    m_k^{\mathrm{II}}(g)=m_k^{\mathrm I}(g),\qquad
    m_k^{\mathrm{III}}(g)=m_k^{\mathrm{IV}}(-g).
    \label{eq:qtm-pair-identities}
\end{equation}

\subsection{The first pair: two independent one-matrix models}

For model I, introduce the real orthogonal change of basis
\begin{equation}
    X=\frac{A+B}{\sqrt2},\qquad Y=\frac{A-B}{\sqrt2}.
\end{equation}
Expanding and using cyclicity of the trace gives
\begin{equation}
    \operatorname{Tr} V(A,B)_{\mathrm I}
    =\operatorname{Tr}\bigl[V(X)+V(Y)\bigr],
    \qquad V(x)=\frac12x^2+\frac g2x^4.
    \label{eq:qtm-factorization}
\end{equation}
This is the standard quartic one-matrix model, whose solution and pure-gravity critical behaviour
are reviewed in \cite{DiFrancesco:1993cyw}. In the conventional potential $x^2/2+\lambda x^4/4$,
the critical coupling is $\lambda_c=-1/12$. Since here $\lambda=2g$, we immediately obtain
$g_c=-\frac1{24}$, and $\gamma_s=-\frac12$. \noindent For completeness, the one-cut resolvent of
$X$ and its endpoint parameter are
\begin{equation}
    W_X(x)  =\frac12\left[x+2gx^3
        -(2gx^2+1+4gs)\sqrt{x^2-4s}\right], \quad s+6gs^2 =1.
    \label{eq:qtm-one-resolvent}
\end{equation}%
At criticality $s=2$, and the density's quadratic prefactor vanishes at each cut endpoint.

The original second moment equals that of $X$, whereas the fourth moment also involves the two
independent factors. Expanding $A=(X+Y)/\sqrt2$ gives
\begin{align}
    m_2^{\mathrm I}=m_2^{\mathrm{II}}
     & =\frac{s(4-s)}3
    =\frac{(1+24g)^{3/2}-1-36g}{216g^2},\label{eq:qtm-first-m2} \\
    m_4^{\mathrm I}=m_4^{\mathrm{II}}
     & =\left[\frac{s(4-s)}3\right]^2+\frac12s^2(3-s)
    =(m_2^{\mathrm I})^2+\frac{1-m_2^{\mathrm I}}{4g}.
    \label{eq:qtm-first-m4}
\end{align}
Expanding perturbatively we find
\begin{align}
    m_2^{\mathrm I,\mathrm{II}}
     & =1-4g+36g^2-432g^3+6048g^4+O(g^5), \\
    m_4^{\mathrm I,\mathrm{II}}
     & =2-17g+196g^2-2664g^3+40176g^4+O(g^5).
    \label{eq:qtm-first-series}
\end{align}

\subsection{The second pair: a standard bilinear two-matrix model}

It suffices to solve model IV. The complex linear transformation
\begin{equation}
    U=\frac{A+iB}{\sqrt2},\qquad
    V=-\frac{A-iB}{\sqrt2},
    \label{eq:qtm-complex-basis}
\end{equation}
gives
\begin{equation}
    \operatorname{Tr} V(A,B)_{\mathrm{IV}}
    =\operatorname{Tr}\bigl[V_1(U)+V_2(V)-UV\bigr],
    \quad V_1(u)=V_2(u)=-\frac g2u^4.
    \label{eq:qtm-bilinear}
\end{equation}
Unlike \eqref{eq:qtm-factorization}, this does not decouple the matrices. It does, however, put the
model in the standard bilinear class studied by Eynard \cite{Eynard:2002kg}. The spectral curve for the resolvent of $U$ takes Eynard's form
\begin{equation}
    E(x,y)=(ax^3-y)(ay^3-x)
    -a^2(x^2y^2-mxy+q)+1=0,
    \label{eq:qtm-second-curve}
\end{equation}%
where $a=-2g$, $m=-\frac1N\bigl\langle\operatorname{Tr}UV\bigr\rangle$,
$q=\frac1N\bigl\langle\operatorname{Tr}U^2V^2\bigr\rangle$, and $W_U(x(z))=V_1'(x(z))-y(z)$. For
cubic $V_1'$ and $V_2'$, the genus-zero parameterisation is
\begin{equation}
    x(z)=\gamma z+\sum_{k=0}^3\alpha_kz^{-k},\qquad
    y(z)=\frac\gamma z+\sum_{k=0}^3\beta_kz^k.
    \label{eq:qtm-general-rational}
\end{equation}%
The symmetry $(U,V)\mapsto(iU,-iV)$ leaves only the coefficients with $k=3$, and exchange symmetry
equates them. Hence
\begin{equation}
    x(z)=\gamma(z+bz^{-3}),\qquad
    y(z)=\gamma(z^{-1}+bz^3).
    \label{eq:qtm-rational}
\end{equation}%

To fix the coefficients, impose $W_U(x)\sim1/x$ at infinity. Cancellation of the $z^3$ term gives
$b=a\gamma^2=-2g\gamma^2$. The remaining leading term is one, giving the normalisation condition
$\gamma^2(3b^2-1)=1$. Setting $r=-\gamma^2$, we obtain
\begin{equation}
    12g^2r^3-r+1=0.
    \label{eq:qtm-r-cubic}
\end{equation}
\noindent It is then straightforward to express the moments in this parameter
\begin{equation}
    m=\frac{1+3r-r^2}{3},\qquad
    q=\frac{1+9r+33r^2-16r^3}{27}.
    \label{eq:qtm-mq-solution}
\end{equation}
\noindent To compute the original moments in terms of $A$ and $B$ we need $h=\frac1N\bigl\langle\operatorname{Tr}U^4\bigr\rangle
    =\frac1N\bigl\langle\operatorname{Tr}V^4\bigr\rangle$ which we can compute by matching the coefficient of $z^{-5}$. This gives
\begin{equation}
    h=2gr^3(2-r).
    \label{eq:qtm-h}
\end{equation}
Using \eqref{eq:qtm-pair-identities}, both models are therefore solved by
\begin{equation}
    \begin{aligned}
        m_2^{\mathrm{III}}(g)=m_2^{\mathrm{IV}}(g)
         & =\frac{1+3r-r^2}{3}, \\
        m_4^{\mathrm{IV}}(g)
         & =\frac{2+9r+15r^2-8r^3}{9}+gr^3(2-r), \\
        m_4^{\mathrm{III}}(g)
         & =\frac{2+9r+15r^2-8r^3}{9}-gr^3(2-r).
    \end{aligned}
    \label{eq:qtm-original-moments}
\end{equation}
\noindent Expanding these expressions perturbatively in $g$ yields
\begin{align}
    m_2^{\mathrm{III},\mathrm{IV}}
     & =1+4g^2+96g^4+3456g^6+O(g^8),\label{eq:qtm-second-series-m2} \\
    m_4^{\mathrm{IV}}
     & =2+g+20g^2+24g^3+576g^4+864g^5+22656g^6+O(g^7), \\
    m_4^{\mathrm{III}}
     & =2-g+20g^2-24g^3+576g^4-864g^5+22656g^6+O(g^7).
    \label{eq:qtm-second-series-m4}
\end{align}

For criticality, recall that a generic finite endpoint $x_e=x(e)$ of the selected resolvent branch
satisfies
\begin{equation}
    x'(e)=0,\qquad x''(e)\neq0,\qquad y'(e)\neq0.
\end{equation}
Expanding $W_U(x(z))=V_1'(x(z))-y(z)$ gives
\begin{equation}
    W_U(x)=W_{\mathrm{reg}}(x)
    -y'(e)\sqrt{\frac{2(x-x_e)}{x''(e)}}
    +O\bigl((x-x_e)^{3/2}\bigr),
    \label{eq:qtm-generic-edge}
\end{equation}
where $W_{\mathrm{reg}}$ is analytic near $x_e$ and the sign of the square root specifies the
sheet. Thus $x'(e)=0$ locates an ordinary endpoint; criticality additionally requires $y'(e)=0$,
eliminating the leading square-root term. Imposing this condition gives the critical point
$g_c=\pm\frac19$. Expanding the correlators about $g_c$, we find $m_2^{\mathrm{sing}}\propto\delta^{3/2}$, where $\delta=1-81g^2$. The corresponding singular planar free energy behaves as $F_0^{\mathrm{sing}}\propto\delta^{5/2}$, giving $\gamma_s=-1/2$.

\bibliographystyle{style/JHEP}
\bibliography{references/references}

\end{document}